\documentclass[preprint]{ptptex}

\usepackage{graphicx}
\newcommand{\be}{\begin{equation}}
\newcommand{\ee}{\end{equation}}
\newcommand{\bea}{\begin{eqnarray}}
\newcommand{\eea}{\end{eqnarray}}
\newcommand{\bean}{\begin{eqnarray*}}
\newcommand{\eean}{\end{eqnarray*}}
\newcommand{\ba}{\begin{array}}
\newcommand{\ea}{\end{array}}

\newcommand{\non}{\nonumber}
\newcommand{\ra}{\rightarrow}

\preprintnumber[3cm]{%<-- [..]: optional width of preprint # column.
hep-ph/0512137\\
December, 2005}
\title{
The Dilepton $Q_T$ Spectrum 
in\\
Transversely Polarized 
Drell-Yan Process in QCD}

\author{Hiroyuki \textsc{Kawamura},$^{1}$
Jiro \textsc{Kodaira},$^{2}$
Hirotaka \textsc{Shimizu},$^{3}$
and\\ 
Kazuhiro \textsc{Tanaka}$^{4}$}
\inst{$^1$Radiation Laboratory, RIKEN, 2-1 Hirosawa, 351-0198, Japan\\
$^2$Theory Division, KEK, Tsukuba, Ibaraki 305-0801, Japan\\
$^3$Dept. of Physics, Hiroshima University, Higashi-Hiroshima 739-8526, Japan\\
$^4$Dept. of Physics, Juntendo University, Inba-gun, Chiba 270-1695, Japan}

\abst{We discuss 
the transverse momentum $Q_T$ distribution of Drell-Yan pair, 
produced in collisions of transversely polarized protons.
We calculate 
the transversely polarized Drell-Yan cross section 
up to 
${\cal O} (\alpha_s )$
in the dimensional regularization,
which gives QCD prediction at large $Q_T$.
For small $Q_T$,
we include all-orders resummation of large logarithms due to emission of  
soft gluons up to next-to-leading 
logarithmic accuracy. At intermediate 
$Q_T$,
the resummation formula is matched with
the fixed-order $\alpha_s$ perturbative results in a systematic way, 
and
we derive the cross section with uniform accuracy over the entire
range of $Q_T$. 
}

\begin{document}

\maketitle

%%%%%%%%%%%%%%%%%%%%%%%%%%%%%%%%%%%%%%%%%%%%%%%%%%%%%%%%%%%%%%%%%%%%%%%
%\section{Introduction}
Hard processes with polarized 
%nucleon 
beams and/or target
enable us to study
spin-dependent dynamics of QCD. 
%and the spin structure of nucleon. 
%By polarized DIS experiments, we measure
%The helicity distributions $\Delta q(x)$ of quarks inside nucleon
%and also  
%estimate $\Delta G(x)$ 
%of gluon from the scaling violations of them.
The helicity distribution $\Delta q(x)$ of quarks inside nucleon 
has been measured in polarized DIS experiments, and also
$\Delta G(x)$ 
of gluon has 
%also 
been 
%roughly 
estimated from the scaling violations of them.
On the other hand, the transversity distribution $\delta q(x)$, 
i.e. the distribution of transversely polarized quarks inside
transversely polarized nucleon, cannot be measured in the inclusive DIS 
due to its chiral-odd nature,\cite{RS:79} 
and remains as the last unknown distribution at the leading twist. 
Transversely polarized Drell-Yan (tDY) process is one of the processes 
where 
%the transversity distribution 
$\delta q(x)$ can be measured,
and has been undertaken at RHIC-Spin experiment.\cite{BSSV:00}

%In view of the limited acceptance in the measurements,
%For comparison with experiment with the limited acceptance of the detectors,
%it is desire
We develop the QCD prediction
of tDY 
%spin-dependent 
cross section,
%\begin{equation}
%\frac{d \sigma}{d Q^2 d Q_T^2 d y d \phi}\ ,
$d \sigma/d Q^2 d Q_T^2 d y d \phi$,
%\label{eq:dc}
%\end{equation}
differential in the transverse momentum $Q_T$ and rapidity $y$  
of the produced lepton pair,
as well as in the dilepton invariant mass $Q$ and in the azimuthal 
angle $\phi$ of one of the leptons with respect to the incoming nucleon's spin axis.
%is required.
Although this $Q_T$- and $y$-differential cross section is fundamental 
in view of comparison with experiment, the corresponding formula has been unknown 
so far
even in the leading order (LO) in QCD:
the lepton-pair production with finite $Q_T$ via the Drell-Yan mechanism has to be accompanied by 
the radiation of at least one recoiling parton, so the LO term
of the cross section is of ${\cal O}(\alpha_s)$.
The corresponding one-loop calculation of the LO term requires
the phase space integration  
separating out the relevant transverse degrees of freedom, 
to extract the $\cos(2\phi)$ dependence
which is characteristic of the spin-dependent cross section of tDY.\cite{RS:79}
In the dimensional regularization, in particular,
the relevant phase space integration in $D$-dimension
is rather cumbersome compared with unpolarized and longitudinally polarized cases.
Furthermore, at small $Q_T$ (``edge regions of the phase
space''), the radiation of soft gluon produces large ``recoil logs'' that spoil
fixed-order perturbation theory, and the corresponding logarithmically enhanced contributions 
have to be resummed to all orders in $\alpha_s$.
Previous works treated {\it $Q_T$- and/or $y$-integrated} tDY cross sections 
{\it to fixed-order} $\alpha_s$, employing the massive gluon scheme\cite{VW:93} 
and the dimensional reduction scheme\cite{CKM:94}
(see also Refs.~\citen{WV:98,MSSV:98}).
In this Letter, we compute the $Q_T$- and $y$-differential cross section of tDY in the LO (${\cal O}(\alpha_s)$) 
%at large $Q_T$ 
in the dimensional regularization scheme. Then the result is consistently combined with
the soft gluon resummation of logarithmically enhanced contributions for small $Q_T$ 
up to next-to-leading logarithmic (NLL) accuracy,
and we derive the first complete result of the well-defined tDY differential cross section 
in the $\overline{\rm MS}$ scheme
for all regions of $Q_T$ at the ``${\rm NLL} + {\rm LO}$'' level.

%QCD corrections to tDY process have been calculated by several authors 
%for $Q_T$-integrated cross sections in various regularization schemes 
%\cite{VW:93,CKM:94,WV:98}, and for $Q_T$-unintegrated cross sections 
%in massive gluon scheme.  
%Here we compute the 1-loop QCD corrections to tDY at a measured $Q_T$ 
%and azimuthal angle $\phi$ of the produced lepton in the dimensional 
%regularization scheme ($\overline{\rm MS}$ scheme).
%For this purpose, the phase space integration in $D$-dimension, 
%separating out the relevant transverse degrees of freedom, 
%is required to extract the $\propto \! \cos(2\phi)$ part of the cross
%section characteristic of the double spin asymmetry of tDY.\cite{RS:79}
%The calculation is rather cumbersome compared with the corresponding 
%calculation in unpolarized and longitudinally polarized cases,
%and has not been performed so far.

%We also include soft gluon effects by all-order resummation of 
%logarithmically enhanced contributions which give the dominant 
%contributions in small $Q_T$ region (``edge regions of the phase
%space'') up to next-to-leading logarithmic (NLL) accuracy.
%Contour deformation prescription is taken to avoid the Landau pole 
%which appears in the b-space (impact parameter space) integration 
%in the all order resummation formula. 

%Finally we present numerical results for $Q_T$-distribution 
%of the spin dependent cross section of tDY.

%%%%%%%%%%%%%%%%%%%%%%%%%%%%%%%%%%%%%%%%%%%%%%%%%%%%%%%%%%%%%%%%%%%%%%%%%%%%%%

%\section{1-loop calculations}

We first consider the tDY cross section to ${\cal O}(\alpha_s)$:
$h_1(P_1,s_1)+h_2(P_2,s_2)\rightarrow l(k_1)+\bar{l}(k_2)+X$,
where $h_1,h_2$ denote nucleons with momentum $P_1,P_2$ and transverse 
spin $s_1,s_2$, and $Q=k_1+k_2$ is the 4-momentum of DY pair.  
Based on QCD factorization, the spin-dependent cross section
$\Delta_T d \sigma \equiv (d \sigma (s_1 , s_2) - d \sigma (s_1 , - s_2))/2$
is given as a convolution,
%\begin{equation}
$\Delta_T d\sigma = \int d x_1 d x_2  
    \delta H (x_1, x_2 ; \mu_F^2)
             \Delta_T d \hat{\sigma} 
%(s_1\,,\,s_2 ; 
(\mu_F^2)$,
%\ , 
%\end{equation}
where $\mu_F$ is the factorization scale, 
%\begin{equation}
$\delta H (x_1 , x_2 ; \mu_F^2) = \sum_i e_i^2 
[\delta q_i(x_1 ; \mu_F^2)\delta \bar{q}_i(x_2 ; \mu_F^2)
+\delta \bar{q}_i(x_1 ; \mu_F^2)\delta q_i(x_2 ; \mu_F^2)]$
%\label{eq:H}
%\end{equation}
is the product of transversity distributions of the two nucleons, and 
$\Delta_T d \hat{\sigma}(\mu_F^2)=(d \hat{\sigma} (s_1 , s_2; \mu_F^2) - d \hat{\sigma} (s_1 , - s_2; \mu_F^2))/2$ 
is the corresponding partonic 
cross section. 
Note that, at the leading twist level, the gluon does not contribute 
to the transversely polarized, chiral-odd process, corresponding to helicity-flip by one unit.
%due to the chiral-odd nature. 
We compute the one-loop corrections to $\Delta_T d \hat{\sigma}(\mu_F^2)$,
which involve the virtual gluon corrections and the real gluon emission 
contributions, e.g.,
$q (p_1 , s_1) + \bar{q} (p_2 , s_2) 
              \to l (k_1) + \bar{l} (k_2) + g$,
with $p_i = x_i P_i$.
We regularize the infrared divergence  
in $D=4 - 2 \epsilon$ dimension, and employ naive anticommuting $\gamma_5$
which is a usual prescription in the transverse spin channel.\cite{WV:98}
In the $\overline{\rm MS}$ scheme, we 
%eventually 
obtain for the differential cross section 
%($S\equiv (P_1 +P_2 )^2$)
% at LO, 
\begin{eqnarray}
\frac{\Delta_T d \sigma}{d Q^2 d Q_T^2 d y d \phi}
= 
%N\, 
\cos{(2 \phi )}
\frac{\alpha^2}{3\, N_c\, S\, Q^2}
 \left[ X\, (Q_T^2 \,,\, Q^2 \,,\, y) 
+ Y\, (Q_T^2 \,,\, Q^2 \,,\, y) \right],
\label{cross section}
\end{eqnarray}
%where 
%$N = \alpha^2 / (3\, N_c\, S\, Q^2)$ with 
with 
$S=(P_1 +P_2 )^2$, and the rapidity
%of the virtual photon 
%is defined as,
%
$y = (1/2) \ln [( Q^0 + Q^3 )/(Q^0 - Q^3)]$
%$y = (1/2) \ln ( Q^+ /Q^- )$
%with 
where $Q^{3}$ is the 
%light-cone 
component 
of the virtual photon momentum $Q^{\mu}$
%the energy and the momentum of the virtual photon 
along the direction
of the colliding beam in the nucleon-nucleon CM system.
%$y$ is the rapidity of virtual photon, and $\phi$ is the azimuthal 
%angle of one of the leptons with respect to the initial spin axis.
%For later convenience, 
We have decomposed the RHS into 
two parts: the function $X$ contains all terms that are singular 
as $Q_T \rightarrow 0$, while $Y$ is of ${\cal O}(\alpha_s)$ and  
finite at $Q_T=0$.
Writing $X = X^{(0)} + X^{(1)}$ as the sum of ${\cal O}(\alpha_s^0)$ and 
${\cal O}(\alpha_s^1)$ contributions,  we have
%\bea
$X^{(0)} = \delta H (x_1^0\,,\,x_2^0\,;\, \mu_F^2 )\ \delta (Q_T^2)$, and
%\label{eq:X0} \\
\bea
X^{(1)} &=& \frac{\alpha_s (\mu_R^2 )}{2 \pi} C_F
       \Biggl\{ \delta H (x_1^0\,,\,x_2^0\,;\, \mu_F^2 ) 
    \left[\, 2 \left( \frac{\ln (Q^2 / Q_T^2 )}{Q_T^2} \right)_+ 
              - \frac{3}{(Q_T^2)_+}
    + \left( \pi^2 - 8  \right) \delta (Q_T^2) \right] 
\nonumber\\
 \!\!\!\!\!\!\!\!\!\!\!\!\!\!\!\!\!+&& \!\!\!\!\!\!\!\!\left(  \frac{1}{(Q_T^2)_+} 
        + \delta (Q_T^2) \ln \frac{Q^2}{\mu_F^2} \right)\!\!\!
       \left[ \int^1_{x_1^0} \frac{d z}{z}
\delta P_{qq}^{(0)} (z)\
        \delta H 
     \left( \frac{x_1^0}{z}, x_2^0 ;\ \mu_F^2 \right)
        +  ( x_1^0 \leftrightarrow x_2^0 ) \right] \Biggr\} ,
\label{eq:X}
\end{eqnarray}
where $\mu_{R}$ is the renormalization scale, $x_1^0 = \sqrt{\tau}\ e^y , x_2^0 =\sqrt{\tau}\ e^{-y}$ 
are the relevant scaling variables with $\tau =Q^2/S$,
and $\delta P_{qq}^{(0)} (z) = 2 z/(1 - z)_+ 
            + (3/2)\, \delta (1 - z)$
is the LO transverse splitting function.\cite{KMHKKV:97,WV:98}
%\cite{AM:90}
The ``+'' distribution to regulate the singularity at $Q_T =0$ is defined 
%with respect to the upper limit $Q^2$ of the $Q_T^2$ integration, 
such that
$\int_0^{p_T^2} d Q_T^2 
     ( [ \ln^n (Q^2 / Q_T^2) ]/Q_T^2 )_+
     = -(1 / [ n + 1 ])  \ln^{n + 1} ( Q^2 / p_T^2 )$.
%
%\begin{equation}
%  \int_0^{p_T^2} d Q_T^2 
%     \left( \frac{\ln^n (Q^2 / Q_T^2)}{Q_T^2} \right)_+
%     = - \frac{1}{n + 1} \ln^{n + 1} \frac{Q^2}{p_T^2} \ .
%\label{eq:plus}
%\end{equation}
%In (\ref{eq:X}), the terms involving $\delta(Q_T^2 )$ come from the 
%virtual gluon corrections, while the other terms 
%The terms with those distributions represent the recoil 
%effects due to the real gluon emissions. 
For the $Y$-term of (\ref{cross section}),
%The finite term $Y$ is obtained by extracting the singular terms from 
%the total ${\cal O}(\alpha_s)$ results:
\bea
%   \lefteqn{
Y
%\, (Q_T^2 \,,\, Q^2 \,,\, y) 
&=& 
         \frac{\alpha_s (\mu_R^2 )}{\pi}\, C_F
\left\{
%\times 
%        \left\{ \, 
         \left[\, \frac{2}{Q^2}\,- \, \frac{3}{Q_T^2}\, 
       \ln \left( 1 + \frac{Q_T^2}{Q^2} \right)  \right] 
%+ \int^1_{\sqrt{\tau_+} \, e^{-y}} \, \frac{d x_2}{x_2 - x_2^+}\,  
%         \delta H (x_1^*\,,\,x_2) \, \frac{\tau}{x_1^* x_2}
%\right. 
%\label{eq:Y}\\
%    && \times 
%\left[ 
            \int^1_{\sqrt{\tau_+} \, e^y} \, \frac{d x_1}{x_1 - x_1^+}\,  
         \delta H (x_1\,,\,x_2^* ; \mu_F^2 ) \, \frac{\tau}{x_1 x_2^*}\right.\non\\
    &+& 
     \left. 
\frac{1}{Q_T^2}\,
      \left[ \int d x_1 \, \delta H_1 
%+ \int d x_2 \, \delta H_2 
               +\frac{1}{2} \delta H (x_1^0\,,\,x_2^0 ; \mu_F^2 )
      \, \ln \frac{(1 - x_1^+)(1 - x_2^+)}{(1 - x_1^0)(1 - x_2^0)} 
           \right]
\right\} + \left(y \rightarrow -y \right)\ , \label{eq:Y}
%\non
%    && \quad + \left. \,  
\eea
where we 
%suppressed the scale dependence, and 
used the shorthand notation for the integral that vanishes for $Q_T =0$, 
\bea
\int dx_1\delta H_1&\equiv&
\int^1_{\sqrt{\tau_+} \, e^y}\ 
      \,  \frac{d x_1}{x_1 - x_1^+} 
        \left[ \delta H (x_1\,,\,x_2^* ;\, \mu_F^2 )
       \, \frac{\tau}{x_1 x_2^*} - \delta H (x_1^0 \,,\,x_2^0 ;\, \mu_F^2 ) \right]\\
   && \hspace{2cm}
    - \int^1_{x_1^0}\ 
      \,  \frac{d x_1}{x_1 - x_1^0} 
        \left[ \delta H (x_1\,,\,x_2^0 ;\, \mu_F^2 )
       \, \frac{x_1^0}{x_1} - \delta H (x_1^0 \,,\,x_2^0 ;\, \mu_F^2 ) \right]\ .
\non
\eea
and defined the variables,
%and $\int dx_2 \delta H_2$ is defined in the same way 
%Kinetic variables in (\ref{eq:Y}) are defined 
%as the 
$\sqrt{\tau_+}=\sqrt{Q_T^2/ S}+\sqrt{\tau+ Q_T^2/ S}$,
$x_{1,2}^{+} = x_{1,2}^{0}\sqrt{1+ Q_T^2 /Q^2 }$,
$x_2^*=( x_1x_2^+-x_1^0x_2^0 )/( x_1-x_1^+ )$,
%
%\begin{equation}
%\sqrt{\tau_+}=\sqrt{\frac{Q_T^2}{S}}+\sqrt{\tau+\frac{Q_T^2}{S}},~~~~
%x_2^*=\frac{x_1x_2^+-x_1^0x_2^0}{x_1-x_1^+},~~~~
%x_{1,2}^{+} = x_{1,2}^{0}\sqrt{1+\frac{Q_T^2}{Q^2}}\ ,
%\label{eq:var}
%\end{equation}
following Ref.~\citen{AEGM:84}.
%\bea
%x_1^+&=&\left(\frac{Q^2+Q_T^2}{S}\right)^{1/2}e^y,~~~~
%x_2^+=\left(\frac{Q^2+Q_T^2}{S}\right)^{1/2}e^{-y},\\
%x_1^*&=&\frac{x_2x_1^+-x_1^0x_2^0}{x_2-x_2^+},~
%x_2^*=\frac{x_1x_2^+-x_1^0x_2^0}{x_1-x_1^+},~
%\sqrt{\tau_+}=\sqrt{\frac{Q_T^2}{S}}+\sqrt{\tau+\frac{Q_T^2}{S}}.
%\eea
Note that $x_{1,2}^+ \rightarrow x_{1,2}^0$ and $x_{2}^* \rightarrow x_{2}^0$
as $Q_T \rightarrow 0$, and
%; it is straightforward to see that
all terms in (\ref{eq:Y}) are finite when $Q_T\ra 0$.
%since 
%$x_{1,2}^*=x_{1,2}^+=x_{1,2}^0$, $\sqrt{\tau_+}=\tau$.

Eq.~(\ref{cross section}) with (\ref{eq:X}) and (\ref{eq:Y}) gives 
the first result for the differential cross section to ${\cal O}(\alpha_s)$ 
in the $\overline{\rm MS}$ scheme.
The result is invariant under the replacement $y \rightarrow -y$ as it should be.
%We note that, integrating (\ref{cross section}) over $Q_T$,
%our result coincides with the corresponding $Q_T$-integrated 
%cross sections obtained in previous work.\cite{WV:98}
We note that, integrating (\ref{cross section}) over $Q_T$,
our result coincides with the corresponding $Q_T$-integrated 
cross sections obtained in previous works employing massive 
gluon scheme\cite{VW:93} and dimensional reduction scheme,\cite{CKM:94} 
via the scheme transformation relation\cite{WV:98} (see also Ref.~\citen{MSSV:98}).
For $Q_T^2 > 0$, $X^{(0)}$ 
%of (\ref{eq:X0}) 
vanishes. Also, the terms proportional 
to $\delta(Q_T^2 )$ in (\ref{eq:X}),
including those associated with the + distribution $( [ \ln^n (Q^2 / Q_T^2) ]/Q_T^2 )_+$,
%(\ref{eq:plus}), 
do not contribute.
The cross section (\ref{cross section}) in this case is of ${\cal O}(\alpha_s )$, and gives the LO QCD
prediction of tDY at the large-$Q_T$ region:
\begin{equation}
\frac{\Delta_T d \sigma^{\rm LO}}{d Q^2 d Q_T^2 d y d \phi}
= 
%N 
\cos{(2 \phi )}
\frac{\alpha^2}{3\, N_c\, S\, Q^2}
 \left[ \left. X^{(1)}(Q_T^2 , Q^2 , y)\right|_{Q_T^2 >0} 
+ Y(Q_T^2 , Q^2 , y) \right].
%\nonumber \\
%&=&
%N
%\cos{(2 \phi )}
%\frac{\alpha_s (\mu_R^2 )}{\pi}\, C_F
%%\left\{
%%\times 
%%        \left\{ \, 
%         \left[ \frac{1}{Q_T^2} + \frac{2}{Q^2}\,- \, \frac{3}{Q_T^2}\, 
%       \ln \left( 1 + \frac{Q_T^2}{Q^2} \right)  \right] 
%%+ \int^1_{\sqrt{\tau_+} \, e^{-y}} \, \frac{d x_2}{x_2 - x_2^+}\,  
%%         \delta H (x_1^*\,,\,x_2) \, \frac{\tau}{x_1^* x_2}
%%\right. 
%%\label{eq:Y}\\
%%    && \times 
%\left[ 
%            \int^1_{\sqrt{\tau_+} \, e^y} \, \frac{d x_1}{x_1 - x_1^+}\,  
%         \delta H (x_1\,,\,x_2^* ; \mu_F^2 ) \, \frac{\tau}{x_1 x_2^*} +(y \rightarrow -y) \right]\ ,
\label{cross section2}
\end{equation}

%reduces to the ordinaly
%functions, 
%$\left( \frac{\ln^n (Q^2 / Q_T^2)}{Q_T^2} \right)_+ \rightarrow \frac{\ln^n (Q^2 / Q_T^2)}{Q_T^2}$.

%%%%%%%%%%%%%%%%%%%%%%%%%%%%%%%%%%%%%%%%%%%%%%%%%%%%%%%%%%%%%%%%%%%%%%%%%%%%%%%%%

%\section{Soft gluon resummation}

The cross section (\ref{cross section}) becomes very large
when $Q_T \ll Q$, due to the terms behaving 
$\sim \alpha_s \ln(Q^2/Q_T^2 )/Q_T^2$ and $\sim \alpha_s /Q_T^2$
in (\ref{eq:X}).
% of the singular part $X$.
%It is well-known that, in unpolarized and longitudinally polarized DY,
Actually, similar large ``recoil logs'' 
%of similar nature 
appear in each order of 
perturbation theory as 
$\alpha_s^n \ln^{2n-1}(Q^2/Q_T^2 )/Q_T^2$, 
$\alpha_s^n \ln^{2n-2}(Q^2/Q_T^2 )/Q_T^2$, and so on,
corresponding to LL, NLL, and higher contributions, respectively,\cite{AEGM:84,CSS:85}
and the resummation of those 
%logarithms 
logarithmically enhanced contributions to 
all orders is necessary to obtain a well-defined, finite 
prediction for the cross section. 
We work out 
%the all-orders resummation of the corresponding 
%logarithmically enhanced contributions
%in (\ref{cross section}) 
this up to the NLL accuracy, 
based on the general formulation\cite{CSS:85} of the $Q_T$ resummation.
This can be 
%conveniently 
carried out in the impact parameter $b$ space,
conjugate to the $Q_T$ space: the singular term $X$ of (\ref{cross section})
%, (\ref{eq:X}) 
is modified 
into the corresponding resummed component, which is expressed as the Fourier 
transform back to the $Q_T$ space. As a result, the first term in the RHS of (\ref{cross section})
is replaced by
\begin{eqnarray}
\frac{\Delta_T d \sigma^{\rm NLL}}{d Q^2 d Q_T^2 d y d \phi}
=
%N 
\cos&& (2 \phi )
\frac{\alpha^2}{3\, N_c\, S\, Q^2}
\sum_{i}e_i^2 \int_0^{\infty} d b \frac{b}{2}
J_0 (b Q_T)\nonumber \\
%\;\;\;\;\;\;\;\;\;\;\;\;\;\;\;
\times
%\!\!\!\!\!\!\!\!
    e^{\, S (b , Q)}  &&
\left[  ( C_{qq} \otimes \delta q_i )
           \left( x_1^0 , \frac{b_0^2}{b^2} \right)      
 ( C_{\bar{q} \bar{q}} \otimes \delta\bar{q}_i )
           \left( x_2^0 , \frac{b_0^2}{b^2} \right)
+ ( x_1^0 \leftrightarrow x_2^0 )\right].
\label{resum}
\end{eqnarray}
Here $J_0 (b Q_T)$ is a Bessel function, 
$b_0 = 2e^{-\gamma_E}$ with $\gamma_E$ the Euler constant,
% is of kinematical origin\cite{CSS:85} ($\gamma_E$ is the Euler constant),
and the large logarithmic corrections are  
resummed into the 
Sudakov factor $e^{S (b , Q)}$ with\\
%\bea
$S(b,Q)=-\int_{b^2_0/b^2}^{Q^2} (d\kappa^2 /\kappa^2 )
%\frac{d\kappa^2}{\kappa^2}
\left\{  A_q (\alpha_s(\kappa^2))\ln( Q^2 / \kappa^2 ) +
B_q(\alpha_s(\kappa^2)) \right\}$.
%\ .
%\label{eq:su}
%\eea
The functions $A_q$, $B_q$ as well as the
coefficient functions $C_{qq}, C_{\bar{q}\bar{q}}$,
are perturbatively calculable: we get
%\bea
$A_q (\alpha_s )= \sum_{n=1}^{\infty} \left( \frac{\alpha_s}{2 \pi} \right)^n A_q^{(n)}$,  
$B_q (\alpha_s )= \sum_{n=1}^{\infty} \left( \frac{\alpha_s}{2 \pi} \right)^n B_q^{(n)}$,
%and
%$C_{ab} (z, \alpha_s )
%%= C_{\bar{q}\bar{q}}
%=\delta_{ab} \, \delta (1 - z) +
%\sum_{n=1}^{\infty} \left( \frac{\alpha_s}{2 \pi}\right)^n C_{ab}^{(n)} (z)$.
%\eea
with
\begin{equation}
A_q^{(1)}=2C_F \ ,~~~~
A_q^{(2)}= 2C_F \left\{\left(\frac{67}{18}-\frac{\pi^2}{6}\right)C_G
-\frac{5}{9}N_f \right\}\ ,~~~~ B_q^{(1)} =-3C_F \ ,\label{eq:AB}\\
\end{equation}
at the present accuracy of NLL, and similarly,
\begin{equation}
C_{qq}(z, \alpha_s )=C_{\bar{q} \bar{q}}(z, \alpha_s )
=\delta(1-z)\left\{1+\frac{\alpha_s}{2\pi}C_F \left(\frac{\pi^2}{2}-4\right) \right\}
+{\cal O}(\alpha_s^2 )\ .
%C_{qq}(z,\alpha_s )=C_{\bar{q}\bar{q}}(z,\alpha_s )=\delta(1-z)C_F\left(\frac{\pi^2}{2}-4\right)\ ,
\label{eq:Cqq}
\end{equation}
Here we have utilized a relation\cite{KT:82} between $A_q$ and 
the DGLAP kernels\cite{KMHKKV:97,WV:98} 
in order to obtain 
%the two-loop term of 
$A_q^{(2)}$ of (\ref{eq:AB}).\cite{Kawamura:2004kj}
The other contributions of (\ref{eq:AB}), (\ref{eq:Cqq}) have been determined so that
the expansion of the above formula (\ref{resum}) in powers
of $\alpha_s (\mu_R^2 )$ reproduces $X$ 
%of (\ref{cross section})
%(\ref{eq:X0}),
with (\ref{eq:X}) to ${\cal O}(\alpha_s)$.
Our results (\ref{eq:AB}) 
%demonstrate
are consistent with the fact 
that the LL ($A_q^{(1)}$) and NLL ($A_q^{(2)},B_q^{(1)}$) 
contributions are universal (process-independent).\cite{DG:00} 
%We note that (\ref{resum}) can be expressed 
%in a form, where the universality of the factorized exponential structure is maximal,~\cite{CAT,BCDeG:03}
%\begin{equation}
%%\frac{\Delta_Td\sigma^{\rm NLL}}{dQ^2dQ_T^2dyd\phi}
%W=\sum_{i}e_i^2 \int_0^{\infty} d b \frac{b}{2}
%J_0 (b Q_T)
%     e^{\, \tilde{S} (b , Q)}  
%( C_{qq} \otimes \delta q_i )
%           \left( x_1^0 , Q^2 \right)      
% ( C_{\bar{q} \bar{q}} \otimes \delta\bar{q}_i )
%           \left( x_2^0 , Q^2 \right) 
%\end{equation}

Eq.~(\ref{resum}) with (\ref{eq:AB}), (\ref{eq:Cqq}) satisfies the evolution 
equation controlled by renormalization
group,\cite{CSS:85}
% equation for the $Q^2$ dependence,\cite{CSS:85} 
and is formally perfect.
As emphasized in Refs.~\citen{LKSV:01,BCDeG:03} for the unpolarized processes, 
%(see also Ref.\cite{LKSV:01}), 
however, several ``reorganization'' in the formula of this type 
%of (\ref{resum})
is necessary for its 
%actual 
consistent evaluation up to required accuracy,
%\cite{LKSV:01,BCDeG:03} 
to treat properly the $b$ integration of (\ref{resum}) 
%from 0 to $\infty$ 
that involves too short as well as long distance.
%compared with our relevant scales.
In the $b$ space, $L\equiv \ln(Q^2 b^2/b_0^2 )$ plays role of the relevant large logarithm, with $b \sim 1/Q_T$.
%Thus it is convenient to 
Expressing the $b$ dependence of the second line in 
%the integrand of 
(\ref{resum})
%has to be systematically organized 
as the corresponding $L$ dependence, and systematically organizing those contributions in 
%the framework of 
the large-logarithmic expansion,
%up to the NLL accuracy,
%, expressing the $b$ dependence of as the corresponding $L$ dependence;
%and subsitutiting the two-loop 
%the second line in the integrand of (\ref{resum})
%the $b$ dependence of
%(\ref{eq:su}) as well as of the parton distributions
%$\delta q_i \left( x_1^0 , b_0^2 / b^2 \right), \delta\bar{q}_i \left( x_2^0 , b_0^2 / b^2 \right)$
%can be expressed as the corresponding $L$ dependence, 
%which .
%ii is known that the $L$ dependent terms can be exponentiated , where
%Based on exponentiation property,
the result is given by the resummation formula with exponentiation of 
the LL terms $\alpha_s^n L^{n+1}$ and the NLL terms $\alpha_s^n L^{n}$ 
in each order of perturbation theory.\cite{LKSV:01,BCDeG:03}
%after exponentiating the all $L$ dependence of those integrands.}
%the $b_0^2 /b^2$ dependence is expressed as $b_0^2 /b^2 =Q^2 e^{-L}$
%is solely through their $L$ dependence in the framework of the resummed large-logarithmic expansion.
%Now the second line of (\ref{resum}) is a function of $L$ without additional dependence on $b$:
%Then, when $L$ becomes large for large $b$, the resummed large logarithms in this resummation formula provide
%the soft gluon effects
%at small $Q_T$ 
%correctly.
%to the NLL accuracy. 
However, $L$ becomes large for small $b$ as well as for large $b$, and thus
the resummation formula contains the unjustified large logarithmic contributions at large $Q_T$. 
To reduce 
%the impact of 
these unjustified contributions, 
%resummed logarithms in the large $Q_T$ region, 
we use a procedure\cite{BCDeG:03} by
performing the replacement
$L \rightarrow \tilde{L} \equiv  \ln(Q^2 b^2/b_0^2 +1 )$;
%in the second line of (\ref{resum});
$\tilde{L}=L+{\cal O}(1/(Qb)^2 )$ for $Qb \gg 1$, 
so $L$ and $\tilde{L}$ are equivalent 
%to arbitrary logarithmic accuracy,i.e., 
to organize the soft gluon resummation at small $Q_T$,
%the replacement does not affect the resummation formula 
%Note that the variables $L$ and $\tilde{L}$ for $Q_T /Q\ll 1$,
but differ at intermediate and large $Q_T$,
%The present choice as $\tilde{L} \rightarrow 0$ for $b\rightarrow 0$
%allows us to kill
%with the unwanted resummed logarithms killed.
%the replacement leads to a different behavior as $\tilde{L} \rightarrow 0$ at large $Q_T$, 
avoiding the unwanted resummed 
contributions as $\tilde{L} \rightarrow 0$ for $Qb \ll 1$.

% at large $Q_T$.

%the contribution of resummed logarithms in (\ref{resum}) 
%$L=\log{(Q^2b^2/b_0^2)} \ra \tilde{L}=\ln{(Q^2b^2/b_0^2+1)} $ in the 
%Sudakov term and the parton distribution functions in (\ref{resum})
%so that the overall normalization is consistent with the fixed order 
%result of the $Q_T$-integrated cross section.\cite{BCDeG:03}    

%One more step is necessary to make the QCD prediction of tDY:
%to properly deal with the $b$-integration of (\ref{resum}),
%which actually involves too short as well as long distance compared with our relevant scales: 
%We also make a further replacement:
The $b$ integration in (\ref{resum})
%the first term of (\ref{eq:cs})
also receives the contribution from long distance:
% (see (\ref{resum})):
similarly to other all-orders resummation formula,
%our result (\ref{resum}) 
the $b$ integration 
%in the long distance
%first term of (\ref{eq:cs})
is suffered from 
%the IR renormalons 
%due to 
the Landau pole at $b= (b_0 /Q)e^{(1/2\beta_0 \alpha_s (Q))}$
due to the Sudakov factor,
%when performing the $b$ integration (see (\ref{resum})), 
and
it is necessary to specify a prescription to deal with
this singularity.\cite{CSS:85,CMNT:96} We follow the method 
introduced in the joint resummation:\cite{LKSV:01}
decomposing the Bessel function of (\ref{resum}) into the two Hankel functions,
%$J_0(bQ_T) = (H_0^{(1)}(bQ_T )+H_0^{(2)}(bQ_T ) )/2$,
we deform the $b$-integration contour of these two terms 
%in the complex $b$ space
%in (\ref{resum}) 
into upper and lower half plane in the complex $b$ space, respectively, and obtain the two convergent
integrals as $|b| \rightarrow \infty$;
%as 
%the two integration branches 
%$b=te^{\pmi\theta}$ 
%($b=te^{-i\theta}$) 
%($0\le t <\infty$, $0< \theta <\pi/4$) to make the integral convergent as $t \rightarrow \infty$.
% in the complex $b$ space for the $H_0^{(1)}$ ($H_0^{(2)}$) part.
%In order to define the b-integration in (\ref{resum}), we need to
%specify a prescription to deal with the Landau pole at 
%$b= (b_0 /Q)e^{(1/2\beta_0\alpha_s (Q))}$ in the Sudakov factor.  
%Here we deform the integration contour in (\ref{resum}) in the complex
%$b$ space, following the method introduced in the joint resummation.
%\cite{LKSV:01}  
%Unlike the frequently used ``$b_{*}$ prescription'' in which 
%the coupling constant is ``frozen'' effectively at
%$b=b_{max}$\cite{CSS:85}, we do not need to introduce an extra parameter
note, this choice of contours 
%in the complex $b$ space
is completely equivalent to the original contour,
%of the first term of (\ref{eq:cs}),
%(\ref{resum}),
order-by-order in $\alpha_s$, when the corresponding formulae 
are expanded in powers
of $\alpha_s$.

%Obviously 
Prescription to define the $b$ integration to avoid the Landau pole
is not unique,
% reflecting IR renormalon ambiguity,
e.g., 
``$b_{*}$ prescription'' to ``freeze'' effectively the $b$ integration  
along the real axis is frequently used.\cite{CSS:85,Kawamura:2004kj}
%The renormalon 
Such ambiguity reflects incompleteness to treat the long-distance contributions
in perturbative framework, and   
should be eventually compensated 
%in the physical quantity 
by 
%the power corrections
%$\sim (b \Lambda_{\rm  QCD})^n$ ($n=2,3, \ldots$) 
%that reflect 
the relevant nonperturbative effects. 
Correspondingly, we make the replacement 
$e^{S (b , Q)}\rightarrow e^{S (b , Q)}F^{NP}(b)$ in (\ref{resum}) 
with the ``minimal'' ansatz for nonperturbative effects\cite{CSS:85,LKSV:01,BCDeG:03} 
with a 
%nonperturbative 
parameter $g_{NP}$,
\begin{equation}
F^{NP}(b)=\exp(-g_{NP} b^2)\ .
\label{eq:np}
\end{equation}

%One price due to the above reorganization: 
%is that 
We note that
the matching of (\ref{resum})
with the fixed-order result $X$,
% ((\ref{eq:X0}),
%(\ref{eq:X})),
% to ${\cal O}(\alpha_s)$, 
which was used to derive (\ref{eq:AB}) and (\ref{eq:Cqq}), 
is now violated at intermediate and large $Q_T$ due to the replacement $L \rightarrow \tilde{L}$.
This can be 
%easily 
%amended 
recovered
by subtracting the 
%corresponding 
${\cal O}(\alpha_s)$ contributions that
violate the matching.\cite{LKSV:01,BCDeG:03} 
Combining the result with the LO cross section (\ref{cross section2}), 
we obtain our differential cross section of tDY at the ``NLL + LO'' level:
\begin{equation}
\frac{\Delta_Td\sigma}{dQ^2dQ_T^2dyd\phi}=
\frac{\Delta_Td\tilde{\sigma}^{\rm NLL}}{dQ^2dQ_T^2dyd\phi}
-\left[ \frac{\Delta_Td \tilde{\sigma}^{\rm NLL}}{dQ^2dQ_T^2dyd\phi}\right]_{{\cal O}(\alpha_s)}
+\frac{\Delta_Td\sigma^{\rm LO}}{dQ^2dQ_T^2dyd\phi}\ ,
\label{eq:cs}
\end{equation}
where $\Delta_Td\tilde{\sigma}^{\rm NLL}/( dQ^2dQ_T^2dyd\phi )$ 
denotes (\ref{resum}) 
with $L \rightarrow \tilde{L}$ and $e^{S (b , Q)}\rightarrow e^{S (b , Q)}F^{NP}(b)$,
%where the first term in R.H.S. is the resummed term given by 
%(\ref{resum}) multiplied by the common factor $Ncos(2\phi)$, and 
and $[\cdots]_{{\cal O}(\alpha_s)}$ denotes 
%the ${\cal O}(\alpha_s)$ in the second term denotes 
%$\left[\Delta_Td\tilde{\sigma}^{\rm NLL}/( dQ^2dQ_T^2dyd\phi )]\right]_{{\cal O}(\alpha_s)}$
the ${\cal O}(\alpha_s)$ terms resulting from the expansion of ``$\cdots$'' in powers of $\alpha_s$
with $g_{NP} \rightarrow 0$.
%to the subtraction term given by the 
%${\cal O}(\alpha_s)$ expansion of the first term. 
%The replacement $L \ra \tilde{L}$ does not affect these terms 
%By construction, 
There is no double counting between the resummed and fixed-order components in (\ref{eq:cs})
%over the whole region of 
for all $Q_T$;
%by combining the resummed term (\ref{resum}) with the leading order 
%result (\ref{cross section}) without double counting, that can  
%be accomplished by the following ``matching'' procedure:    
e.g., at 
%small 
$Q_T \ll Q$, both the second and third terms in (\ref{eq:cs}) become 
%very 
large 
%and so the singular terms as $Q_T\ra 0$ 
%in the second and the third terms 
but cancel with each other. In particular, the integral of (\ref{eq:cs}) over $Q_T$ reproduces
that of (\ref{cross section}) {\it exactly}, because\cite{BCDeG:03} 
$\tilde{L}= 0$ at $b=0$; our $Q_T$ resummation does not affect the total cross section.
%to ${\cal O}(\alpha_s )$. 
Eq.~(\ref{eq:cs}) gives the well-defined tDY differential cross section 
in the $\overline{\rm MS}$ scheme
%, with uniform accuracy 
over the entire
range of $Q_T$.

%The NLO parton distributions in the $\overline{\rm MS}$ scheme
%have to be used in the above formula.

%%%%%%%%%%%%%%%%%%%%%%%%%%%%%%%%%%%%%%%%%%%%%%%%%%%%%%%%%%%%%%%%%%%%%%%%%%%%%%%%%%%%

%\section{The $Q_T$ spectrum of Drell-Yan dimuon production at RHIC}

As an application, 
we compute (\ref{eq:cs}) as a function of $Q_T$
%multiplied by $2Q_T$
for $\sqrt{S}=200$ GeV, $Q=8$ GeV, $y=2$, and $\phi=0$,
% in Fig.1,
which corresponds to 
%QCD prediction for 
the tDY $Q_T$-spectrum 
for the detection of dimuons with the PHENIX detector at RHIC.
%For applications of our result (\ref{eq:cs}) to phenomenology,
%For this purpose,
As the first estimate of this quantity,
we use
%need 
the following nonperturbative inputs, for which our knowledge is uncertain: 
%we use a model\cite{MSSV:98} 
%or which our present knowledge is restricted.
for the transversity $\delta q(x, \mu^2)$ in (\ref{cross section2}), (\ref{resum}),
%we use a model,\cite{MSSV:98} 
%in which we
%guided by the Soffer inequality, 
%$2\delta | q(x,\mu^2) | \le q(x,\mu^2)+\Delta q(x,\mu^2)$,
%where $q$ and $\Delta q$ are the unpolarized and helicity distributions:
we saturate the ``Soffer bound'' as $\delta q(x,\mu_0^2)=[q(x,\mu_0^2)+\Delta q(x,\mu_0^2)]/2$
at low input scale $\mu_0 \simeq 0.6$ GeV with 
%using 
the NLO density and helicity distributions
%GRV and GRSV 
%densities 
$q(x,\mu_0^2)$ and $\Delta q(x,\mu_0^2)$, 
and evolve $\delta q(x, \mu_0^2)$ to higher $\mu^2$ with the NLO DGLAP 
kernel,\cite{KMHKKV:97,WV:98} see Ref.\citen{MSSV:98} for the detail. 
As for 
the nonperturbative parameter 
$g_{NP}$ of (\ref{eq:np}),
we use 
%the value 
$g_{NP}\simeq0.5$ GeV$^2$ 
suggested by the study of Ref.\citen{KS}.
%a simplest form 
%$F^{NP}(b)=\exp{(-g_{NP}b^2)}$ with a non-perturbative parameter:$g_{NP}$.   
%These inputs are to be determined from a global fit of various observables.  
%the short-dashed, 
The solid curve in Fig.1
%and long-dashed lines 
shows (\ref{eq:cs}), multiplied by $2Q_T$,
with $g_{NP}=0.5$ GeV$^2$.
%$g_{NP}=0.3$ GeV$^2$, 0.5 GeV$^2$
%, and 0.8 GeV$^2$, respectively.
We also show the LO result using (\ref{cross section2}) by the dashed curve, and the contribution from
the $Y$-term of (\ref{eq:Y}) by the dotted curve. 
For calculation of all curves, we chose 
%the factorization and renormalization scales as 
$\mu_F = \mu_R =Q$.
%As an application, 
%we compute (\ref{eq:cs}) 
%multiplied by $2Q_T$
%for $\sqrt{S}=200$ GeV, $Q=8$ GeV, $y=2$, and $\phi=0$ in Fig.1,
%which corresponds to the QCD prediction for the $Q_T$ spectrum of the tDY 
%for the detection of dimuons with the PHENIX detector at RHIC.
%to be observed at the forward muon detector 
%at RHIC PHENIX experiment.\cite{BSSV:00}
%Fig.1 shows the $Q_T$ spectrum of the dilepton pair in tDY 
%for $\sqrt{S}=200$ GeV, $Q=8$ GeV, $y=2$ and $\phi=0$,  
%corresponding to di-muon detection at the forward muon detector 
%at RHIC PHENIX experiment.\cite{BSSV:00} 
%Here we take a model for transversity $\delta q(x)$ given by 
%Martin et al.\cite{MSSV:98} in which Soffer's inequality is saturated 
%at the initial scale: 
%$2\delta q(x,Q_0^2)=q(x,Q_0^2)+\Delta q(x,Q_0^2)$
%at $Q_0\sim 0.8 GeV$ (``maximal scenario''). 
%As for non-perturbative 
%function which corresponds to intrinsic transverse momentum distribution
%of partons inside a hadron, we take a simplest form 
%$F^{NP}(b)=\exp{(-g_{NP}b^2)}$ with a non-perturbative parameter:$g_{NP}$.   
%These inputs are to be determined from a global fit of various observables. 
%The dot-dashed line in Fig.1 shows the leading order (LO) cross section  
%given by (\ref{cross section}). 
%The short-dashed, solid, long-dashed lines show the cross section 
%given by NLL resummation matched with LO cross section with 
%non-perturbative parameter $g_{NP}=0.3GeV^2,0.5GeV^2, 0.8GeV^2$ respectively. 
%As expected, 
The LO result becomes large and diverges as $Q_T \rightarrow 0$,
while the ``NLL + LO'' result is finite and well-behaved over all regions of $Q_T$. 
%Comparing the ``NLL + LO'' results with the LO result and the $Y$-term contribution,
%In particular,
The soft gluon resummation gives dominant effects around the peak of the solid curve,
i.e., 
%not only in the
%small $Q_T$ or around the ``peak'' region, but also 
at intermediate $Q_T$ as well as small $Q_T$.
%region higher than the ``peak'' region.
%To demonstrate role of nonperturbative effects (\ref{eq:np}), 
In Fig.1, 
the results of (\ref{eq:cs}) with
$g_{NP}=0.3$ GeV$^2$ and 0.8 GeV$^2$ are also shown 
by the dot-dashed and two-dot-dashed lines, respectively:
%These results demonstrate that the $Q_T$ spectrum is sensitive to the value of $g_{NP}$: 
the larger $g_{NP}$ gives the broader spectrum with the higher peak position, because
%naively 
the larger $g_{NP}$ of (\ref{eq:np}) corresponds to the larger ``intrinsic transverse momentum'' 
of partons inside nucleon. 
%We note that 
The impact of the nonperturbative effects (\ref{eq:np})
becomes milder when the short-distance contributions are more dominated, i.e., 
for the case with larger $Q^2$ (see also Refs.\citen{LKSV:01,BCDeG:03}).
 
In conclusion the perturbative effects relevant for the dilepton $Q_T$ spectrum in the tDY in QCD
are now under control over the entire range of $Q_T$.
%in this energy region unlike the case of Higgs production.\cite{BCDeG:03} 
%The dotted line shows Y-term in (\ref{eq:Y}) which is sub-dominant in
%small $Q_T$ region but become important in large $Q_T$ region. 
Apparently further systematic studies in various
kinematic region of $pp$ collision, as well as of $p\bar{p}$ collision\cite{Shimizu:2005fp},
with the resummation formalism
%, including the threshold resummation\cite{Shimizu:2005fp},
are indispensable to reveal $\delta q(x)$ and also the intrinsic transverse motion of quarks.
%and will be presented  
%in the forthcoming paper.
%\cite{KKST} 

\vspace{-0.1cm}

\begin{figure}
\centerline{\includegraphics[width=7cm]{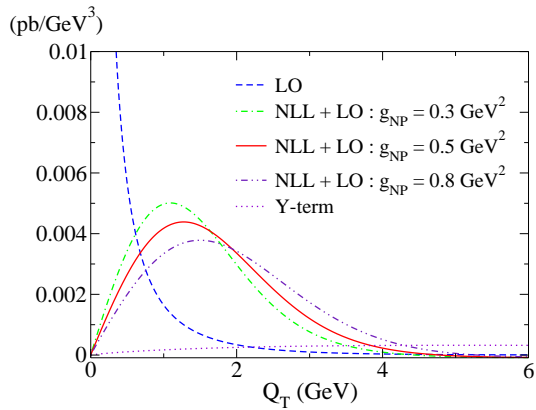}}
\caption{$Q_T$ distribution $\Delta_T d \sigma/d Q^2 d Q_T d y d \phi$
at $\sqrt{S}=200$ GeV, $Q=8$ GeV, $y=2$ and $\phi=0$. }
\end{figure}

%\vspace{-0.4cm}

%\section*{Acknowledgements}
We would like to thank Werner Vogelsang for providing us with 
a fortran code of transversity distributions and valuable 
discussions. 
We also thank Stefano Catani for fruitful discussions. 
The work of J.K. 
%was supported by the Grant-in-Aid for 
%Scientific Research No. C-16540255. The work of 
and K.T. was 
supported by the Grant-in-Aid for Scientific Research Nos. C-16540255 and C-16540266.

\end{document}